\documentclass[authoryear,preprint,review,12pt]{elsarticle}

\usepackage{graphics}

\usepackage{amssymb}
\usepackage{bm}
\usepackage{natbib}

\newcommand{\pp}[1]{\phantom{#1}}
\newcommand{\be}{\begin{eqnarray}}
\newcommand{\ee}{\end{eqnarray}}

\newcommand{\Tr}{{\,\rm Tr\,}}

\newcommand{\ba}{\begin{array}}
\newcommand{\ea}{\end{array}}

\journal{Ecological Modeling}

\begin{document}

\begin{frontmatter}



\title{Systems, environments, and soliton rate equations: A non-Kolmogorovian framework for population dynamics}


\author[DA]{Diederik Aerts}\author[DA,MC]{Marek Czachor}\author[MK]{Maciej Kuna}\author[DA]{Sandro Sozzo}
\address[DA]{Centrum Leo Apostel (CLEA) and Foundations of the Exact Sciences (FUND),
Vrije Universiteit Brussel, 1050 Brussels, Belgium}
\address[MC]{Katedra Fizyki Teoretycznej i Informatyki Kwantowej,
Politechnika Gda\'nska, 80-952 Gda\'nsk, Poland\footnote{Fax: (+48 58) 347-28-21, Phone: (+48 58) 347 2225, e-mail: mczachor@pg.gda.pl}}
\address[MK]{Katedra Rachunku Prawdopodobieñswa i Biomatematyki,
Politechnika Gda\'nska, 80-952 Gda\'nsk, Poland}
\begin{abstract}
Soliton rate equations are based on non-Kolmogorovian models of probability and naturally include autocatalytic processes. The formalism is not widely known but has great unexplored potential for applications to systems interacting with environments. Beginning with links of contextuality to non-Kolmogorovity we introduce the general formalism of soliton rate equations and work out explicit examples of subsystems interacting with environments. Of particular interest is the case of a soliton autocatalytic rate equation coupled to a linear conservative environment, a formal way of expressing seasonal changes. Depending on strength of the system-environment coupling we observe phenomena analogous to hibernation or even complete blocking of decay of a population.
\end{abstract}

\begin{keyword}
rate equations \sep soliton dynamics \sep non-Kolmogorovian probability \sep biodiversity



\end{keyword}

\end{frontmatter}



\section{Introduction}

It is quite typical that soliton equations discovered in one domain of science turn out to have applications in completely different fields. The so-called nonlinear Schr\"odinger equation is a perfect example. Its applications range from surface waves on deep water \citep{Zakharov} to nonlinear effects in DNA \citep{Bishop} and fiber optics \citep{Kibler}. The name comes from mathematical similarity to the basic equation of quantum mechanics. However, this is only a similarity. The true Schr\"odinger equation is always linear.

In this paper we concentrate on another soliton system, Darboux-integrable von Neumann equations \citep{Leble Czachor(1998),UCKL,CCU}, whose origin goes back to studies on generalizations of quantum mechanics \citep{Czachor(1997)}, but which has never gained great popularity among physicists. The reasons are similar to those from our previous example: von Neumann equations occurring in quantum mechanics are always linear, but Darboux-integrable von Neumann dynamics can be nonlinear. It naturally includes catalytic and autocatalytic processes \citep{Aerts Czachor(2006),Aerts et al.(2006)}, but the probability calculus behind it is contextual and hence non-Kolmogorovian. So, is it possible that the formalism discovered in attempts of generalizing quantum mechanics has unexpected applications in a different domain? If so, what kind of criteria should one use to identify these new applications? The hints come from autocatalysis and probability.

Contextuality plays in probability a role similar to that of curvature in geometry. In geometry curvature measures non-commutativity of translations. In probability contextuality measures non-commutativity of questions. In geometry curvature requires local charts: taken together they form a global object, an atlas, covering the entire curved space. The charts are pairwise compatible so that one can explore the whole curved object by thumbing the atlas. The whole global structure is called a manifold. In probability, questions asked earlier define contexts for those that are asked later. Typically, a triple of questions $A\to B\to C$ has to be described in two steps, $A\to B$ and then $B\to C$, and each step involves a different probability space. The global structure is not a probability space in the sense of \cite{Kolmogorov(1956)}. Rather, it is an object of a manifold type \citep{Gudder}.

Non-Kolmogorovian probabilities are sometimes termed
quantum probabilities, but the name is misleading and its origins are historic and not ontological. Non-Kolmogorovian structures are as ubiquitous as non-Euclidean geometries \citep{Khrennikov(2010)}. Non-Euclidean geometries were discovered before the advent of general relativity, otherwise one would speak of ``gravitational geometry". Non-Kolmogorovian probability was discovered in quantum mechanics, but after some 30 years of studies in logical foundations of quantum mechanics it has become clear that non-Kolmogorovity has nothing to do with microphysics
\citep{AccardiFedullo1982,Aerts1986,Pitowsky1989}. It is thus striking that the search for fundamental laws of ecology has led ecologists to probabilistic structures of a propensity type \citep{Ulanowicz(1999),Ulanowicz(2009)}, which are known to be conceptually close to probability models of quantum mechanics \citep{Popper}. However, as opposed to some other authors, we do not attempt to identify concrete physical media (electromagnetic fields \citep{Tiezzi1}, liquid water \citep{Tiezzi2}) that might be responsible for physical origins of ecosystem dynamics. Contextuality is enough to generate non-Kolmogorovity.

The goal of the present paper is to introduce formal mathematical structures that are needed in discussion of nonlinear models based on non-Kolmogorovian probability. In Section 2 we begin our analysis with a simple illustration of non-Kolmogorovity: A cyclic competition. In Section 3 we discuss a model of non-Kolmogorovian probability (based on vectors), and further generalize it in Section 4 to the density-matrix formalism. In Section 5 the density-matrix
formalism is used to reformulate (possibly nonlinear) rate equations in a Lax-von Neumann form. In order to develop some intuitions we first concentrate on a linear example. In Section 6 we introduce the notion of a hierarchy of coupled environments. The first examples of nonlinear rate equations and their Lax-von Neumann representation occur in Section 7. Section 8 introduces the main subject of this paper: soliton rate equations. An equation describing a linear environment coupled to a nonlinear subsystem is explicitly analyzed. Choosing various values of parameters we plot evolution of populations for shorter and longer time scales and alternative couplings between systems and their environment. All these examples are based on exact solutions of the associated system of nonlinear rate equations. Section 9 is devoted to the specific case of a periodically changing environment. Finally, in Section 10, we give a preliminary analysis of soliton systems with dissipation and the role of environment in possible blocking or hibernation of dissipative decay of populations. Section 11 is the first step that goes beyond the soliton dynamics. Here we introduce a new generalization of replicator equations applicable to games where players try to modify the rules of the game during its course. The equation is of a von Neumann form and seems to have applications to real-life evolutionary games, a subject that will be discussed in more detail in a future work, cf. \citep{Aerts et al.(2013)}.

\section{Cyclic competition is non-Kolmogorovian}

A simple form of cyclic competition occurs in the rock-paper-scissors game: Rock $R$ destroys scissors $S$, scissors destroy paper $P$, paper destroys rock. In a kinetic form the game corresponds to
\be
R+S &\to& R+\textrm{decay products},\label{R+S}\\
S+P &\to& S+\textrm{decay products},\label{S+P}\\
P+R &\to& P+\textrm{decay products}.\label{P+R}
\ee
Logical and probabilistic version of the game is illustrated by Fig.~1.
There are two players and three random variables $S$, $P$, $R$ with binary values 0, 1. Players choose which random variables will be measured, and then appropriate pairs of binary values occur with probabilities $p(S=0,R=1)=p(P=0,S=1)=p(R=0,P=1)=1$. The remaining probabilities vanish.

In spite of trivial statistics the problem is formally quite subtle, a fact mentioned in the context of ``non-quantum" probability already by \cite{Vorobev}. Vorobev's ideas led mathematicians to the notion of a contextual marginal problem \citep{Fritz}, i.e.  the question if and when a collection of probabilities can be regarded as a result of computing marginal probabilities for pairs from joint probabilities of triples. If triple joint probabilities do not exist, one has to resort to local probability spaces, and one arrives at a manifold-like non-Kolmogorovian structure.

In fact, we can model the game on three independent probability spaces corresponding to the three pairs of random variables shown in Fig.~1a. Each such pair can be realized in an experiment. The resulting data can be manipulated in standard ways, but one has to be cautious not to mix data from different {\it alternative\/} experiments. To give an example, the average of $S$ can be computed by means of
\be
\overline{S}
=
\sum_{a,b=0}^1 ap(S=a,P=b)=p(S=1,P=0)=1.\nonumber
\ee
If one computes the average as follows,
\be
\overline{S}
=
\sum_{a,b=0}^1 ap(S=a,R=b)=0,\nonumber
\ee
a contradictory result is obtained. This does not mean that $0=1$. There is no contradiction if one takes into account that logically it is not allowed to treat $S$ in the $S$-$P$ experiment as the same $S$ as in the $S$-$R$ one. $R$ and $P$ form different contexts for $S$. In quantum mechanical terminology one can say that the two players constitute a non-local system. One can also reformulate the RPS game in a way that explicitly turns it into a variant of Einstein-Podolsky-Rosen experiment \citep{evolution2011} where Bell's inequality \citep{bell1964} is violated.

But then we arrive at another difficulty. The rock-paper-scissors game is a canonical example of cyclic competition. It has its analogues in population dynamics of plankton, lizards or bacteria. Standard approaches to kinetic dynamics imply that elementary processes (\ref{R+S})-(\ref{P+R}) lead to nonlinear rate equations of the form
\be
{[\dot S]}&=& -k_S [S][R]+\dots,\label{[S+R]}\\
{[\dot R]}&=& -k_R [R][P]+\dots,\label{[R+P]}\\
{[\dot P]}&=& -k_P [P][S]+\dots.\label{[P+S]}
\ee
The question is if $[S]$ in (\ref{[S+R]}) is the same $[S]$ as the one in (\ref{[P+S]})? If so, why? Clearly, (\ref{[S+R]}) represents a process where $S$ interacts with $R$, while (\ref{[P+S]}) occurs due to interactions between $P$ and $S$. But we have just shown that a naive mixing of two the two $S$s leads to contradictions.

So the difficulty is that we have to combine simultaneously two apparently contradictory aspects. On the one hand, in the RPS game the other player chooses either $P$ or $R$, but $P$ and $R$ cannot be chosen simultaneously. The two contexts for $S$ cannot occur at the same instant of time. On the other hand, however, $[S]$ in both equations is a time dependent variable $[S](t)$, with the same values of $t$ in (\ref{[S+R]}) and  (\ref{[P+S]}). The problem is typical for quantum mechanical evolutions. Its solution is known since 1920s, when it was understood how to probabilistically interpret solutions of the Schr\"odinger equation.

\section{Non-Kolmogorovian probability}

Kolmogorovian model of probability leads to conceptual difficulties whenever order of ``questions" (i.e. context) is not irrelevant. The Kolmogorovian algorithm for constructing conditional and joint probabilities is formally based on projections on subsets of events, $\chi_AX=A\subset X$. In practical applications $\chi_A$ is often represented by characteristic functions
\be
\chi_A(x)
&=&
\left\{
\begin{array}{cl}
1 & \textrm{if }x\in A\\
0 & \textrm{if }x\not\in A\\
\end{array}
\right.
\ee
If $\mu$ is a probability measure on the set of events $X$,  $\mu(X)=1$, then probability of finding $x\in A\subset X$ is defined as
\be
p(A)
&=&
\mu(\chi_AX).
\ee

In case of joint probabilities
\be
p(A\cap B)
&=&
\mu(\chi_{A\cap B}X)=\mu(\chi_{A}\chi_{B}X)
\nonumber\\
&=&
\mu(\chi_{B}\chi_{A}X)=p(B\cap A).
\ee
Conditional probabilities are defined via the Bayes rule
\be
p(A\cap B) &=& p(A)p(B|A),\\
p(B|A)
&=&
\frac{p(A\cap B)}{p(A)}=
\frac{\mu(\chi_{A}\chi_{B}X)}{\mu(\chi_{A}X)}.
\ee
Based on set-theoretic properties one can derive a number of inequalities that have to be satisfied by $p(A)$. For example, $0\leq p(A)\leq 1$, $p(A\cap B) \leq p(A)$, or
\be
p(A\cap B\cap C) = p(A\cap C\cap B) \leq p(A\cap B).\label{in1}
\ee
One of the simplest problems with (\ref{in1}) occurs in quantum mechanics with the so-called Malus law stating that joint probability of photon's passage through two polarizers $A$ and $B$ is
\be
p(A\cap B)=p(A)p(B|A)=\frac{1}{2}\cos^2\phi
\ee
where $\phi$ is the angle between the polarizers and $p(A)=1/2$ is the probability of transmission through the first polarizer.
Obviously, if $A$ and $B$ are perpendicular then $p(A\cap B)=0$ since $p(B|A)=0$. But for polarizers tilted by 45 degrees one finds
$p(B|A)=1/2$. In consequence, if one takes a third polarizer $C$ tilted by 45 degrees with respect to mutually perpendicular $A$ and $B$, then
\be
p(A\cap C\cap B)&=&p(A)p(C|A)p(B|C)\nonumber\\
&=&
\frac{1}{2}\cdot\frac{1}{2}\cdot\frac{1}{2}=\frac{1}{8},\\
p(A\cap B\cap C)&=&p(A)p(B|A)p(C|B)\nonumber\\
&=&\frac{1}{2}\cdot 0\cdot\frac{1}{2}=0,\\
p(A\cap B)&=&0,
\ee
leading to the counterintuitive inequality
\be
p(A\cap C\cap B)=\frac{1}{8} > 0=p(A\cap B).
\ee
By $A\cap C\cap B$ we denote an event where the photon first crosses $A$, then $C$, and finally $B$. From a logical point of view the interpretation makes sense. Indeed, if the source of light is to the left of $A$ and the detector is to the right of $B$, then the act of detection means that the particle had to be transmitted through all the polarizers: It had to pass through $A$, and $C$, and $B$. It follows that the detection event can be regarded as $A\cap C\cap B$.

Symbolically, for photons and polarizers $A\to C\to B$ may be allowed, whereas the direct process $A\to B$ may be forbidden. This is the essence of the problem and not the fact that we speak of quantum particles.
Analogous diagrams are typical of chemistry, biology, psychology, cognitive science...

In such theories, if $p(A\cap C\cap B)\neq p(A\cap B\cap C)$ then one cannot assume that
\be
p(A\cap B\cap C)
=
\mu(\chi_{A\cap B\cap C}X)=\mu(\chi_{A}\chi_{B}\chi_{C}X).
\ee
One needs a different mathematical model.
In some (but not all) generalized models of probability one begins with the basic property of ``questions" (or ``propositions" in the logic parlance),
\be
\chi_{A\cap A}=\chi_{A}=\chi_{A}\chi_{A},
\ee
representing the fact that $\chi_{A}$ is a {\it projection\/}. Non-commutativity of propositions can be then trivially obtained if one notices that typical projectors of {\it vectors\/} do not commute (even in the 2D real plane the only projections that commute are those on parallel or perpendicular directions).

So let now $X$ be a vector and $P_A$ a projector on direction $A$. $P_A$ may be imagined as a matrix satisfying $P_A^2=P_A$. Let $\langle X|Y\rangle$ be a scalar product of two vectors. Take a unit $X$ and define $\mu(X)=\langle X|X\rangle=1$, and
\be
p(A)=\langle P_AX|P_AX\rangle=\langle X|P_A|X\rangle.
\ee
$\langle X|P_A|X\rangle$ is a matrix element of $P_A$. If $\sum_A P_A=\mathbb{I}$ (the identity matrix) then
\be
\sum_Ap(A)=\langle X|\sum_AP_A|X\rangle=1.
\ee
The set of projectors that are {\it complete\/}, i.e. that sum to $\mathbb{I}$, represents {\it a\/} maximal collection of simultaneously ``askable" questions. The associated probabilities sum to unity. However, even in the 2D plane there are infinitely many such sets. It follows that there may be infinitely many complete maximal sets of simultaneously meaningful questions/propositions, but questions belonging to two different complete sets cannot be {\it simultaneously\/} considered (but can be asked one after another).

The rule for joint and conditional probabilities becomes (if ``first $A$ then $B$" are ``asked")
\be
p(A\cap B)
&=&
\langle P_AX|P_AX\rangle
\frac{\langle P_BP_AX|P_BP_AX\rangle}{\langle P_AX|P_AX\rangle}
\nonumber\\
&=&
\langle X|P_AP_BP_A|X\rangle=\langle X|E_{A\cap B}|X\rangle.\label{ABA}
\ee
Probability in general depends on the order of questions,
\be
p(B\cap A)
=
\langle X|P_BP_AP_B|X\rangle=\langle X|E_{B\cap A}|X\rangle.\label{BAB}
\ee
In symbolic notation this means that $A\to B$ occurs with different probability than $B\to A$, a generic property of chemical or biological processes.
Formula (\ref{ABA}) represents probability of the answer ``yes" to the question $B$, if it is asked in the {\it context\/} of a positive answer to an earlier question $A$. Note that $E_{B\cap A}$ and $E_{A\cap B}$ are projectors only if $P_AP_B= P_BP_A$. A general $E_{B\cap A}$ is the so called positive operator valued measure, POVM \citep{POVM}. A POVM that is not a projector can be regarded as a non-Kolmogorovian analogue of a fuzzy-set membership function.

The Kolmogorovian formula can be written as a commutative version of the same rule
\be
p(A\cap B)
&=&
\mu(\chi_{A}X)\frac{\mu(\chi_{B}\chi_{A}X)}{\mu(\chi_{A}X)}
=
\mu(\chi_{A}\chi_{B}\chi_{A}X)\nonumber\\
&=&\mu(\chi_{B}\chi_{A}\chi_{B}X)
=p(B\cap A).\nonumber
\ee
What we have sketched is just an {\it example\/} of a non-Kolmogorovian model. Another model, more directly applicable to population dynamics, is described in the next section.

\section{Density operator model of probability}

Let us employ Dirac's notation where complex column vectors $X$ are denoted by $|X\rangle$. The scalar product
\be
\langle Y|X\rangle
&=&
\sum_{j=1}^n \overline{Y_j}X_j
=
(\overline{Y_1},\dots,\overline{Y_n})
\left(
\begin{array}{c}
X_1\\
\vdots\\
X_n
\end{array}
\right)\label{YX}
\ee
can be regarded as a matrix multiplication of the $1\times n$ matrix
\be
\langle Y|=(\overline{Y_1},\dots,\overline{Y_n})=\big(|Y\rangle\big)^\dag
\ee
times the $n\times 1$ matrix $|X\rangle$ ($\dag$ denotes Hermitian conjugation). (\ref{YX}) regarded as a matrix product is, strictly speaking, not a number but a complex $1\times 1$ matrix $\big(|Y\rangle\big)^\dag |X\rangle=\langle Y||X\rangle$. Dirac's notation is based on identification of  $1\times 1$ matrices with complex numbers, i.e. $\langle Y||X\rangle=\langle Y|X\rangle$. Now we can treat the formula
$p(A)=\langle X|P_A|X\rangle$ as the product of three matrices: $1\times n$ $\langle X|$, $n\times n$ $P_A$, and $n\times 1$ $|X\rangle$.
The density operator formalism naturally occurs if one introduces the $n\times n$ matrix
\be
\rho_X = |X\rangle\langle X|
&=&
\left(
\begin{array}{c}
X_1\\
\vdots\\
X_n
\end{array}
\right)
(\overline{X_1},\dots,\overline{X_n})
\nonumber\\
&=&
\left(
\begin{array}{ccc}
|X_1|^2 & \dots & X_1\overline{X_n}\\
\vdots & \ddots & \vdots\\
X_n\overline{X_1}& \dots & |X_n|^2
\end{array}
\right)
\label{XX}
\ee
satisfying
\be
p(A)
&=&\langle X|P_A|X\rangle\nonumber\\
&=&\Tr \big(P_A \rho_X\big) \quad\textrm{(definition)}\\
\Tr\rho_X &=& \langle X|X\rangle=1\quad\textrm{(normalization)}\\
\langle Y|\rho_X|Y\rangle
&=&
|\langle Y|X\rangle|^2\geq 0\quad\textrm{(positivity)}\\
\rho_X^\dag &=& \rho_X\quad\textrm{(reality)}.
\ee
Normalization, positivity and Hermiticity are characteristic of any combination $\rho=\sum_X\lambda_X\rho_X$ if $\lambda_X$ are probabilities, i.e. nonnegative real numbers that sum to 1. Joint and conditional probabilities for a general $\rho$ are defined by
\be
p(A\cap B) &=& \Tr(P_AP_BP_A\rho)\nonumber\\
&=&
\underbrace{\Tr(P_A\rho)}_{p(A)}\underbrace{\frac{\Tr(P_BP_A\rho P_A)}{\Tr(P_A\rho P_A)}}_{p(B|A)}.
\ee
Hermiticity of $\rho$ implies that its eigenvalues are real. Positivity means that these eigenvalues are nonnegative, and normalization guarantees that they sum to 1. If one skips normalization but keeps Hermiticity and positivity then for a projector $P$ the number
$\Tr(P\rho)$ is nonnegative. If $\rho(t)$ is a positive Hermitian solution of some differential equation, then $\Tr\big(P\rho(t)\big)$ is a nonnegative (kinetic) variable.

Let us finally show that a single density operator $\rho$ encodes in an extremely efficient way a number of kinetic variables \citep{Aerts Czachor(2006)}.
$\rho$ is an operator that acts in a linear (Hilbert) space $\cal H$ whose basis is given by a set
$\{|n\rangle, \,\langle n|m\rangle=\delta_{nm}\}$ (there are infinitely many such bases).
Matrix elements of $\rho$ are in general complex
\be
\rho_{nm}&=&\langle n|\rho|m\rangle=x_{nm}+iy_{nm}\\
\overline{\rho_{nm}}&=&x_{nm}-iy_{nm}=x_{mn}+iy_{mn}=\rho_{mn}
\ee
and thus $x_{mn}=x_{nm}$, $y_{mn}=-y_{nm}$, $y_{nn}=0$.

The diagonal elements $\rho_{nn}=x_{nn}$ are themselves probabilities since
\be
\rho_{nn}&=&\langle n|\rho|n\rangle=\Tr P_n\rho=p_n=x_{nn}
\ee
where $P_n=|n\rangle\langle n|$. Now, let
\be
|jk\rangle &=& \frac{1}{\sqrt{2}}\big(|j\rangle+|k\rangle\big),\\
|jk'\rangle &=& \frac{1}{\sqrt{2}}\big(|j\rangle-i|k\rangle\big),
\ee
$P_{jk}=|jk\rangle\langle jk|$, $P'_{jk}=|jk'\rangle\langle jk'|$. Then
\be
x_{jk}
&=&
p_{jk}-\frac{1}{2}p_{j}-\frac{1}{2}p_{k},\\
y_{jk}
&=&
p'_{jk}-\frac{1}{2}p_{j}-\frac{1}{2}p_{k},\\
\rho_{jk}
&=&
p_{jk}+ip'_{jk}-\frac{e^{i\pi/4}}{\sqrt{2}}(p_j+p_k),\label{p+ip'}
\ee
where $p_{jk}=\Tr P_{jk} \rho$, $p'_{jk}=\Tr P'_{jk} \rho$.
It follows that a single $\rho$ encodes three families of probabilities:
$\{p_n\}$, $\{p_{jk}\}$, and $\{p'_{jk}\}$. They are associated with three families of projectors: $\{P_n\}$, $\{P_{jk}\}$, and $\{P'_{jk}\}$. Additional relations between the probabilities follow from
\be
x_{jj}= p_j =p_{jj}-p_{j},\quad
y_{jj}
=
0=
p'_{jj}-p_{j}
\ee
and the resulting formula
$
p_j=p'_{jj}=p_{jj}/2.
$
Let us note that $\{P_n\}$ is complete, i.e. $\sum_{n=1}^{\dim \cal H}P_n=\mathbb{I}$,
$\sum_{n=1}^{\dim \cal H}p_n=1$. In order to understand completness properties of
$\{P_{jk}\}$ and $\{P'_{jk}\}$ we introduce, for $j<k$, two additional types of vectors and their associated projectors:
\be
|jk^\perp\rangle &=& \frac{1}{\sqrt{2}}\big(|j\rangle-|k\rangle\big),\\
|jk'{}^\perp\rangle &=&\frac{1}{\sqrt{2}}\big(|j\rangle+i|k\rangle\big),
\ee
$P_{jk}^\perp=|jk^\perp\rangle\langle jk^\perp|$,
$P_{jk}'{}\!\!^{\perp}=|jk'{}^\perp\rangle\langle jk'{}^\perp|$.
The completeness relations for $\{P_{jk}\}$ and $\{P'_{jk}\}$ follow from the formula
\be
P_{jk}+P_{jk}^\perp&=& P_j+P_k=P'_{jk}+P_{jk}'{}\!\!^{\perp}.
\ee
For any three Hermitian matrices satisfying $[A,B]=iC$ one can prove the  standard-deviation uncertainty relation
$
\Delta A \Delta B\geq \frac{1}{2}|\langle C\rangle|
$
where $\Delta A =\sqrt{\langle A^2\rangle-\langle A\rangle^2}$, $\langle A\rangle=
\Tr (A\rho)$, etc. If $A=P=P^2$ is a one-dimensional projector then $\langle P\rangle=p$ is a probability and one finds $\Delta P=\sqrt{p(1-p)}$.
Two propositions $P_1$ and $P_2$ are {\it complementary\/} if  $\Delta P_1\Delta P_2
\geq \varepsilon>0$. Complementarity means that reducing error $\Delta P_1$ we inevitably increase the one for $\Delta P_2$.

In order to show that
${P}_{j}$ and ${P}_{jk}$ are complementary we compute
\be
[{P}_{j},{P}_{jk}]
&=&
\textstyle{\frac{1}{2}}\big(|j\rangle\langle k|
-
|k\rangle\langle j|\big)
=iC
\ee
and
$\langle C\rangle=y_{kj}$,
\be
\sqrt{p_{j}(1-p_{j})}\sqrt{p_{jk}(1-p_{jk})}
\geq \textstyle{\frac{1}{2}}|y_{kj}|.
\ee
The variable $y_{kj}$ measures complementarity of $P_{j}$ and $P_{jk}$.

The probabilities inherent in a single $\rho$ have been so far defined with respect to a fixed basis $|n\rangle$. Being arbitrary, the basis could be replaced by {\it any\/} other orthonormal basis $|\tilde n\rangle$. Repeating the construction we would then arrive at new sets of projectors,
$\tilde P_n=|\tilde n\rangle\langle\tilde n|$, and so on. They would lead to new families of probabilities, $\tilde p_n=\Tr \tilde P_n\rho$, etc. There is no a priori rule that privileges one basis, so all these probabilities can be meaningful. What is important, all $P_n$, $P_{nm}$, $P'_{nm}$, may be complementary to all $\tilde P_n$, $\tilde P_{nm}$, $\tilde P'_{nm}$. What this practically means is that performing measurements of, say, the probability $p_{12}$, we inevitably influence possible future results of the remaining complementary probabilities. Measurement of $p_{12}$ creates a {\it nontrivial context\/} for measurements of many (perhaps even all) probabilities $\tilde p_n$, $\tilde p_{nm}$, $\tilde p'_{nm}$, as well as for $p'_{12}$, $p_{1}$, and $p_{2}$.

Finally, let us make the trivial remark that the standard Kolmogorovian model is a special case of the density operator formalism. The corresponding $\rho$ is then a combination
\be
\rho=\sum_{j=1}^n p_j P_j
\ee
where $P_j$ belong to the {\it same\/} maximal set. This is equivalent to restricting $\rho$ to diagonal matrices
\be
\rho
=
\left(
\begin{array}{ccc}
p_1 & & 0\\
& \ddots & \\
0& & p_n
\end{array}
\right)
\ee
with (Kolmogorovian) probabilities on the diagonal.

Nonlinear rate equations of a generalized Lotka-Volterra type can be formulated in terms of $\rho(t)$. What is interesting, the formalism involving probabilities encoded by means of $\rho(t)$ automatically allows us to formulate nonlinear rate equations in the so-called von Neumann, Liouville-von Neumann, or Lax forms. The latter property turns out to be essential for soliton structures and thus opens a possibility of solving very complicated coupled systems of rate equations by soliton techniques. In soliton and non-Kolmogorovian contexts it is most appropriate to speak of Lax-von Neumann forms of rate equations: ``Lax", since it stresses the soliton aspect, and ``von Neumann" since the model of probability is based on density operators.

\subsection{Formal definition of contextuality}

Consider two random variables $A$ and $B$ such that the joint probability $p(A=a\cap B=b)$ is defined for all values $a$ and $b$ of $A$ and $B$, respectively. We assume that measurements of $A$ are performed first, i.e. $A$ is a context for $B$.
Conditional and joint probabilities are defined (in both Kolmogorovian and non-Kolmogorovian frameworks) by the Bayes rule
\be
p(A=a\cap B=b)=p(B=b|A=a)p(A=a).
\ee
W define the probability of $B=b$ in the context of $A$ by
\be
p(B=b|A)=\sum_a p(A=a\cap B=b)
\ee
The model (or problem) is non-contextual if
\be
p(B=b|A_1)=p(B=b|A_2)=p(B=b)
\ee
for all random variables $A_1$ and $A_2$. In the density-matrix projector model the rule reads
\be
p(B=b|A)=\sum_a \Tr(\rho P_{A=a}P_{B=b}P_{A=a}),
\ee
where $\sum_a P_{A=a}=\sum_b P_{B=b}=\mathbb{I}$.
If $P_{A=a}P_{B=b}=P_{B=b}P_{A=a}$ then
\be
p(B=b|A) &=&\sum_a \Tr(\rho P_{A=a}P_{B=b}P_{A=a})\nonumber\\
&=&\sum_a \Tr(\rho P_{A=a}P_{B=b})=\Tr(\rho P_{B=b})\nonumber\\
&=&p(B=b).
\ee
So, sensitivity to order of questions indicates contextuality.
In \citep{Aerts et al.(2013)} we show that contextuality in this sense is generic for nontrivial evolutionary games.
Kolmogorovian models, with characteristic functions in the role of projectors, are non-contextual. Practical applications may require POVMs
$E_{A=a\cap B=b}$ yet more general than $P_{A=a}P_{B=b}P_{A=a}$. This type of generalization is employed in \citep{Aerts et al.(2013)} in order to reconstruct practically observed probabilities occurring in the RPS game played by {\it Uta stansburiana\/} lizards.

\section{Preliminaries on rate equations in Lax-von Neumann form}

Returning to the RPS game we can now solve the paradox. Namely, the dynamical aspect is localized in $\rho(t)$ which is a matrix collecting all the possible propensities in all the possible contexts. In order to choose which context we need for $S$, say, we define two  POVMs $E_{P\cap S}$ and $E_{R\cap S}$, so that
\be
p_{R\cap S}(t) &=& \Tr\rho(t)E_{R\cap S},\\
p_{P\cap S}(t) &=& \Tr\rho(t)E_{P\cap S}.
\ee
The dynamics of probabilities is defined through the dynamics of $\rho(t)$. Contextuality is present if $\sum_{p=0}^1E_{P=p\cap S=s}\neq
\sum_{r=0}^1E_{R=r\cap S=s}$. In Kolmogorovian probability the latter would be impossible since the characteristic functions $\chi_{P\cap S}$ and $\chi_{R\cap S}$ always do commute.
It remains to define the dynamics of $\rho(t)$.

A Lax-von Neumann form of rate equations is
\be
\dot\rho = -i[H(\rho),\rho].\label{vN}
\ee
The dot denotes time derivative, $[X,Y]=XY-YX$, and $H(\rho)$ is a linear Hermitian operator that depends on $\rho$. We assume that $H(\rho)=H(\rho)^\dag$ if $\rho=\rho^\dag$. If $\rho(t)=r(\rho_0,t)$ is a concrete solution of (\ref{vN}) with initial condition $\rho(0)=\rho_0$, then
\be
H\big(\rho(t)\big)
=
H\big(r(\rho_0,t)\big)=h(\rho_0,t)
\ee
is a time-dependent operator satisfying, for this particular $\rho(t)=r(\rho_0,t)$, the linear equation
\be
\dot\rho(t) = -i[h(\rho_0,t),\rho(t)].\label{vN'}
\ee
It is then known that there exists a unitary operator $U(\rho_0,t)$ satisfying
\be
\rho(t)
=
U(\rho_0,t)\rho(0) U(\rho_0,t)^\dag.\label{UrU}
\ee
The nonlinearity of (\ref{vN}) is reflected in the dependence of $U(\rho_0,t)$ on the initial condition $\rho_0$.
If $U$ does not depend on $\rho_0$, i.e. is the same for all initial conditions then the dynamics $t\mapsto\rho(t)$ is linear.
The form of solution (\ref{UrU}) is very important since it shows that $\rho(t)$ and $\rho(0)$ are related by unitary equivalence. Two unitarily equivalent Hermitian matrices have the same eigenvalues. In consequence, $\rho(t)$ is positive whenever $\rho(0)$ is positive. This, finally, guarantees that in order to guarantee positivity of $\Tr P\rho(t)$ it is enough to start with a positive initial condition $\rho(0)$.

The central issue of the paper is the soliton, hence nonlinear dynamics of populations. However, since nonlinearities are not here exactly of the usual type, an interpretation in terms of abundances of populations vs. amounts of resources requires some new intuitions that are easier to develop on examples of linear evolutions. So let us begin with more detailed analysis of linear rate equations.

\subsection{Linear systems}
\label{Sec-linear}

Let us now explicitly show a simple $2\times 2$ example of standard rate equations associated with their linear Lax-von Neumann form. In previous sections we were not careful enough to distinguish between linear operators and their matrices, but now we need more precision. The same operator may be represented by different matrices, sice the notion of a matrix is basis dependent. So let $h$ denote the matrix of the numbers $h_{mn}=\langle m|H|n\rangle$, where $|n\rangle$ are arbitrarily chosen orthonormal basis vectors. Similarly, let $\varrho$ be the matrix consisting of the numbers $\varrho_{mn}=\langle m|\rho|n\rangle$.

Now consider
\be
h
&=&
\left(
\begin{array}{cc}
k_{1} & k_{12}+ik'_{12}\\
k_{12}-ik'_{12} & k_{2}
\end{array}
\right),\label{Hgen}\\
\varrho
&=&
\left(
\begin{array}{cc}
\varrho_{11} & \varrho_{12}\\
\varrho_{21} & \varrho_{22}
\end{array}
\right)
.
\ee
The Lax-von Neumann equation
\be
i\dot \varrho=[h,\varrho]
\ee
is equivalent to four rate equations,
\be
\dot p_1
&=&
\left(k_{12}-k'_{12}\right) (p_1+p_2)
\nonumber\\
&\pp=&
+2 k'_{12} p_{12}-2 k_{12} p'_{12},\\
\dot p_2
&=&
\left(k'_{12}-k_{12}\right) (p_1+p_2)\nonumber\\
&\pp=&
-2 k'_{12} p_{12}+2 k_{12} p'_{12},\\
\dot p_{12}
&=&
\left(k_1-k_2\right) p'_{12}
+\left(-\frac{k_1-k_2}{2}-k'_{12}\right) p_1
\nonumber\\
&\pp=&
+\left(-\frac{k_1-k_2}{2}+k'_{12}\right) p_2,\\
\dot p'_{12}
&=&
\left(\frac{k_1-k_2}{2}+k_{12}\right) p_1+\left(\frac{k_1-k_2}{2}-k_{12}\right) p_2
\nonumber\\
&\pp=&
+\left(k_2-k_1\right) p_{12}.
\ee
It is clear that $h$ is a matrix encoding kinetic constants of the dynamics. In order to identify appropriate types of interactions between the populations we have to know concrete values of the $k$s occurring in $h$.

So assume, for example, that
\be
k_1=1,\,k_2=2k_1,\,k_{12}=3k_1,\,k'_{12}=4k_1, \label{4k}
\ee
and denote $p_1=A$, $p_2=B$, $p_{12}=C$, $p'_{12}=D$.
Then
\be
\dot A
&=&
-(A+B)+8 C-6 D,\label{dotA}\\
\dot B
&=&
(A+B)-8 C+6 D,\label{dotB}\\
\dot C
&=&
-\frac{7}{2} A+\frac{9}{2} B- D
,\label{dotC}\\
\dot D
&=&
\frac{5}{2} A-\frac{7}{2} B+ C.\label{dotD}
\ee
We know by construction that $A(t)$, $B(t)$, $C(t)$, $D(t)$ are nonnegative if $A(0)$, $B(0)$, $C(0)$, $D(0)$ are nonnegative, so this is a kinetic system, although not exactly of a chemical type. Adding the first two equations we find that $A+B=\Tr\varrho$ is time independent (one of the conservation laws typical of Lax-von Neumann dynamics).

Switching to another basis $|\tilde n\rangle$ we will obtain a different set of rate equations. For example, the eigenvectors, $H|\tilde n\rangle=h_n|\tilde n\rangle$, make $\tilde h_{mn}=\langle \tilde m|H|\tilde n\rangle$ diagonal,
\be
\tilde h
&=&
\left(
\begin{array}{cc}
h_1 & 0\\
0 & h_2
\end{array}
\right).\label{H}
\ee
The corresponding probabilities
\be
\tilde p_{n}=\tilde\varrho_{nn}=\langle \tilde n|\rho|\tilde n\rangle=\Tr \tilde P_n\rho
=
\Tr \big(|\tilde n\rangle\langle \tilde n|\rho\big)
\ee
etc., satisfy
\be
\dot {\tilde p}{}_{12}
&=&
k \Big({\tilde p}{}_{12}'-\frac{{\tilde p}{}_1 +   {\tilde p}{}_2}{2}\Big),\\
\dot {\tilde p}{}_{12}'
&=&
-k \Big({\tilde p}{}_{12}-\frac{{\tilde p}{}_1 +   {\tilde p}{}_2}{2}\Big),\\
\dot {\tilde p}{}_{1} &=& \dot {\tilde p}{}_{2}\,=\,0,
\ee
with $k=h_1-h_2$.
A general solution of these equations,
\be
{\tilde p}{}_{12}'(t) &=& \frac{{\tilde p}{}_1+{\tilde p}{}_2}{2}
\nonumber\\
&\pp=&
+
\Big({\tilde p}{}_{12}'(0)-\frac{{\tilde p}{}_1+{\tilde p}{}_2}{2}\Big)\cos kt
\nonumber\\
&\pp=&
-
\Big({\tilde p}{}_{12}(0)-\frac{{\tilde p}{}_1+{\tilde p}{}_2}{2}\Big)\sin kt,\label{p12'}\\
{\tilde p}{}_{12}(t) &=& \frac{{\tilde p}{}_1+{\tilde p}{}_2}{2}
\nonumber\\
&\pp=&
+
\Big({\tilde p}{}_{12}'(0)-\frac{{\tilde p}{}_1+{\tilde p}{}_2}{2}\Big)\sin kt
\nonumber\\
&\pp=&
+
\Big({\tilde p}{}_{12}(0)-\frac{{\tilde p}{}_1+{\tilde p}{}_2}{2}\Big)\cos kt,\label{p12}\\
{\tilde p}{}_1(t) &=& {\tilde p}{}_1(0)\,=\,{\tilde p}{}_1,\label{p1}\\
{\tilde p}{}_2(t) &=& {\tilde p}{}_2(0)\,=\,{\tilde p}{}_2,\label{p2}
\ee
with nonnegative initial condition  at $t=0$, remains nonnegative for all $t\neq 0$.
The sum of all the probabilities is not time independent since they do not belong to the same single maximal complete set.

If we take the parameters (\ref{4k}), and denote $\tilde A=\tilde p_1$, $\tilde B=\tilde p_2$, $\tilde C=\tilde p_{12}$, $\tilde D=\tilde p'_{12}$,
then
\be
\dot {\tilde A} &=& \dot {\tilde B}\,=\,0,\label{tildeA+B}\\
\dot {\tilde C}
&=&
-\sqrt{101} \Big({\tilde D}-\frac{{\tilde A} +   {\tilde B}}{2}\Big),\label{tildeC}\\
\dot {\tilde D}
&=&
\sqrt{101} \Big({\tilde C}-\frac{{\tilde A} +   {\tilde B}}{2}\Big),\label{tildeD}\\
{\tilde A}+{\tilde B} &=& A+B.\label{A+B}
\ee
As we can see practically all elements of the kinetic system, including kinetic constants, have changed.

The non-Kolmogorovity of the probability model allows for coexistence of (\ref{dotA})--(\ref{dotD}) and (\ref{tildeA+B})--(\ref{tildeD}) as representing context-dependent aspects of {\it the same\/} dynamical system. One can show that there exists a time-independent linear invertible transformation that maps $\tilde A$, $\tilde B$, $\tilde C$, $\tilde D$ into $A$, $B$, $C$, $D$. In terminology of information theory such a map is called a lossless communication channel. These statements will hold true also for solutions of nonlinear equations discussed later on in this paper.

\subsection{Interpretation of the linear model}

All the dynamical variables are nonnegative so can be interpreted in terms of population abundances.
The constant of motion $(A+B)/2=(\tilde A+\tilde B)/2$ defines a threshold that changes signs of derivatives of $\tilde C(t)$ and $\tilde D(t)$.
Population $\tilde D(t)$ grows as long as the abundance $\tilde C(t)$ is greater than $(A+B)/2$. When $\tilde D(t)$ becomes greater than $(A+B)/2$, the population represented by $\tilde C(t)$ starts to decrease. At the moment the abundance $\tilde C(t)$ becomes smaller than the threshold value $(A+B)/2$, the growth of $\tilde D(t)$ stops, and a decay begins.

Let us finally make the initial condition concrete. Assume $\tilde A=\tilde B=1/2$, $\tilde C(0)=1$, $\tilde D(0)=1/2$, corresponding to
\be
\tilde\varrho(0)
=
\frac{1}{2}
\left(
\begin{array}{cc}
1 & 1\\
1 & 1
\end{array}
\right)
\ee
whose eigenvalues are 0 and 1. Since $\tilde\varrho(0)$ is a positive operator (its eigenvalues are nonnegative) then for any projector $P$ one finds $\Tr \big(P\varrho(t)\big)\geq 0$. In particular
\be
\Tr \big(\tilde P_{12}\varrho(t)\big)
&=&
\tilde C(t)
=
\frac{1}{2}
+
\frac{1}{2}\cos kt,\label{p1,12}\\
\Tr \big(\tilde P_{12}'\varrho(t)\big)
&=&
\tilde D(t) = \frac{1}{2}
-
\frac{1}{2}\sin kt
.
\ee
$\tilde C(t)$ and $\tilde D(t)$ are complementary (belong to different complete sets) because $[\tilde P_{12},\tilde P'_{12}]\neq 0$, and thus do not have to sum to 1.
They are shifted in phase with respect to each other in analogy to typical predator--prey abundances following from Lotka--Volterra models.

\section{Dynamics of subsystems in varying environments}

Distinction between subsystems and their environments can be formalized in analogy to what one does in physics of open systems: State of a subsystem is influenced by the state of its environment, but not vice versa. Subsystems are coupled to environments in a non-symmetric way, a fact expressing the intuition that environments are ``large" as compared to their inhabitants. If one cannot neglect the influence of a subsystem on its environment, then the whole ``subsystem plus its environment" has to be considered as a single system, so that separation into two distinguished parts is no longer meaningful.

Let us now consider the following coupled set of rate equations
\be
\dot\varrho_1 &=& F_1(\varrho_1),\\
\dot\varrho_2 &=& F_2(\varrho_1,\varrho_2),\\
\dot\varrho_3 &=& F_3(\varrho_1,\varrho_2,\varrho_3),\\
&\vdots&
\ee
Some of them may be representable in a Lax-von Neumann form, some of them perhaps not.
The system described by $\varrho_1$ plays a role of environment for the remaining subsystems. The one described by $\varrho_2$ is the environment for $\varrho_3$, $\varrho_4$, and so on. The collection of rate equations has to be solved in a hierarchical way. One begins with $\varrho_1$ since the associated differential equation is closed. Once one finds a given $\varrho_1(t)=r(t)$, one switches to
\be
\dot\varrho_2 &=& F_2(r,\varrho_2).
\ee
Technically speaking, what one has to solve will be a set of coupled nonlinear rate equations with time dependent coefficients.
At a first glance the problem looks, in its generality, hopelessly difficult. However, we will see that the power of soliton Lax-von Neumann equations may allow us to find explicit, exact, and highly nontrivial special solutions for the whole hierarchies of environments.

\section{Nonlinear systems}

Linear Lax-von Neumann equations lead to linear systems of rate equations. Nonlinear rate equations of a generalized Lotka-Volterra type will occur if one takes less trivial $H(\varrho)$. For example, for $H(\varrho)=A\varrho+\varrho A$ one finds
\be
i\dot \varrho
&=&
[A\varrho+\varrho A,\varrho]=[A,\varrho^2].\label{r^2}
\ee
In the two-dimensional case, the nonlinearity in (\ref{r^2}) is yet ``too weak" since all nonlinear terms cancel out in the corresponding rate equations (this does not happen if $A$ is a $3\times 3$ matrix). (\ref{r^2}) is a particular case of
\be
\dot\varrho=-i[H,f(\varrho)]=-i[H_f(\varrho),\varrho]\label{fvN}
\ee
where $f$ is an arbitrary function and $H$ is a linear operator. Given $f$ one can find $H_f(\varrho)$, so this is indeed a Lax-von Neumann equation in the sense we have discussed above. The fact that (\ref{fvN}) is a soliton rate system was established by \cite{UCKL}.

Yet another class of non-linear rate equations is obtained if one takes
\be
K
&=&
\left(
\begin{array}{cc}
0 & k\\
k & 0
\end{array}
\right),
\ee
and replaces (\ref{H}) by
\be
H(\varrho) &=&
\left(
\begin{array}{cc}
\Tr (K\varrho)& 0\\
0 & 0
\end{array}
\right)
=
\left(
\begin{array}{cc}
2 kx_{12}& 0\\
0 & 0
\end{array}
\right)\nonumber\\
&=&
k\left(
\begin{array}{cc}
2p_{12}-p_1-p_2 & 0\\
0 & 0
\end{array}
\right)
\ee
The Lax-von Neumann equation
\be
i\dot \varrho=[H(\varrho),\varrho]
\ee
becomes equivalent to a system of coupled catalytic/auto-catalytic rate equations,
\be
\dot p_{12}
&=&
2k \Big(p_{12}-\frac{p_1 +   p_2}{2}\Big)\Big(p_{12}'-\frac{p_1 +   p_2}{2}\Big),\\
\dot p_{12}'
&=&
-2k \Big(p_{12}-\frac{p_1 +   p_2}{2}\Big)^2,\\
\dot p_{1} &=& \dot p_{2}\,=\,0.
\ee

\section{Soliton rate equations}

Soliton rate equations are not widely known even in the community of soliton-oriented mathematicians and physicists. An interested reader should probably begin with some general introduction to Darboux transformations \citep{MaS}, and then with a more specialized literature such as the monograph by \cite{Leble}. Alternatively, one can directly start with the original papers, beginning with the first work of \cite{Leble Czachor(1998)} where a soliton technique of integrating (\ref{r^2}) was introduced. Generalization (\ref{fvN}) of (\ref{r^2}) is at the top of an infinite hierarchy of more complicated equations systematically cataloged by \cite{CCU}. Let us stress that $H$ may be a differential operator and $\varrho$ could be infinitely dimensional. If one relaxes Hermiticity constraints then soliton Lax-von Neumann equations turn out to contain a large variety of integrable lattice systems.

The term ``soliton" is understood here in the general sense of ``those equations that are solvable by soliton methods". In technical terms this means that there exist Darboux-B\"acklund-covariant Lax pairs whose compatibility conditions are identical to the soliton system in question.

Let us illustrate the latter statement by the soliton system (\ref{fvN}). The Lax pair can be given here in various forms, but the following one is sufficent for our purposes,
\be
z_\mu\varphi_\mu
&=&
(\varrho -\mu H)\varphi_\mu,\label{2-a}\\
i\dot\varphi_\mu
&=&
\frac{1}{\mu}f(\varrho)\varphi_\mu.\label{2-b}
\ee
Here $\mu$, $z_\mu$ are (time independent) complex numbers and $\varphi_\mu$ is a matrix (it can be a vector, i.e. a 1-column matrix). $\varphi_\mu$ can exist if
\be
i\dot{(z_\mu\varphi_\mu)}
&=&
(i\dot\varrho -\mu i\dot H)\varphi_\mu
+
(\varrho -\mu H)\frac{1}{\mu}f(\varrho)\varphi_\mu\nonumber
\ee
is equivalent to
\be
z_\mu i\dot\varphi_\mu
&=&
\frac{1}{\mu}f(\varrho)(\varrho -\mu H)\varphi_\mu.\nonumber
\ee
Since $\varrho f(\varrho)=f(\varrho)\varrho$, the condition reduces to
\be
\big(i\dot\varrho -[H,f(\varrho)]\big)\varphi_\mu=i\mu \dot H\varphi_\mu
\ee
If $\dot H=0$ and $i\dot\varrho -[H,f(\varrho)]=0$, which is our Lax-von Neumann system, then $\varphi_\mu$ may exist.

The crucial step is given by the following
{\it theorem on Darboux covariance of the Lax pair\/}: Let $\varphi_\mu(t)$ be a 1-column matrix which is a solution of the Lax pair for some $\varrho(t)$ satisfying the compatibility condition (\ref{fvN}). Then
\be
\varrho[1]
&=&
\varrho+(\mu-\bar\mu)[P,H]\\
&=&
\Big(\mathbb{I}+\frac{\mu-\bar\mu}{\bar\mu}P\Big)\varrho \Big(\mathbb{I}+\frac{\bar\mu-\mu}{\mu}P\Big),\\
P &=& \varphi_\mu \varphi_\mu^\dag/(\varphi_\mu^\dag\varphi_\mu),
\ee
is also a solution of (\ref{fvN}).

The theorem allows us to find a new solution $\varrho[1]$ given some known solution $\varrho$. But how to find a $\varrho$? $\varrho$ cannot be too simple, say $\varrho=\mathbb{I}$, or a more general $\varrho$ but commuting with $H$, since then $\varrho[1]=\varrho$. Various tricks leading to nontrivial $\varrho[1]$ have been nevertheless invented. For example, if $f(\varrho)=\varrho^2$ one finds
\be
[H,\varrho^2]=[H\varrho+\varrho H,\varrho].
\ee
A time independent $\varrho$ that anticommutes with $H$, $\varrho H=-H\varrho$, leads to $P$ that does not commute with $H$, so that $\varrho[1]\neq \varrho$.
The procedure can be iterated: $\varrho=\varrho[0]\to \varrho[1]\to\varrho[2]\to\dots$. In soliton terminology $\varrho[1]$ is a 1-soliton solution, since it is derived by means of a single Darboux-B\"acklund transformation, $\varrho[0]\to \varrho[1]$.

I should be stressed that although dimensions of the matrices considered in our examples will be small, the method works in any dimension (even infinite). This is why Lax-von Neumann rate equations are naturally suited for problems involving multiple competition of a large number of species. If continuous time is replaced by a discrete time-step, the Lax-von Neumann system turns into a kind of intransitive network whose vertices are defined by matrix elements of $\rho$. Intransitive networks have recently been applied to the problem of biodiversity by \cite{AllesinaLevine2011}.

So let us return to the problem of systems interacting with environments, and let
\be
\dot \varrho_1=F_1(\varrho_1)
\ee
be any system of linear or nonlinear rate equations that describe the environment.
In order to specify the form of $F_2$ in
\be
\dot\varrho_2=F_2(\varrho_1,\varrho_2)
\ee
let us assume that in the absence of environment this is a soliton rate system involving (auto-)catalytic processes of order not higher than 2. Assuming that system--environment interaction also involves only elementary second-order processes we should finally obtain rate equations of order not higher than 3. A simple model possessing these characteristics is
\be
i\dot\varrho_2
=
\big(\alpha+\beta \Tr (P_{12}\varrho_1)\big)[A,(1-s)\varrho_2+s\varrho_2^2],\label{F2}
\ee
where $\alpha,\beta,s\in \mathbb{R}$. Probabilities $p_{1,12}=\Tr (P_{12}\varrho_1)$ are derived from the solution $\varrho_1(t)$. $A$ is a $n\times n$ Hermitian matrix ($n=3$ is the lowest dimension for which quadratic terms do not cancel out in resulting rate equations). Parameter $s$ allows us to control additional interactions between populations that were evolving independently of one another in the linear case discussed in Sec. \ref{Sec-linear}.

We begin with the observation that if we know a solution $w$ of
\be
i\dot w
&=&
[A,(1-s)w+sw^2].\label{F20}
\ee
then
\be
\varrho(t)=w\big(\alpha t+\textstyle{\beta\int_0^td\tau\,p_{1,12}(\tau)}\big)\label{rho_2}
\ee
solves (\ref{F2}). $w(t)$ can be found by soliton techniques introduced by \cite{Leble Czachor(1998)}.

A relatively simple particular 1-soliton solution $w=\varrho[1]$ of (\ref{F20}) can be found if
\be
A=
\left(
\begin{array}{ccc}
0 & 0 & 0\\
0 & 1 & 0\\
0 & 0 & 2
\end{array}
\right)
\ee
and we reduce the number of independent variables by imposing the constraints $w_{12}=w_{23}$, $w_{11}=w_{33}$. Then \citep{Leble Czachor(1998)}
\be
w(t)
=
\frac{1}{15+\sqrt{5}}
\left(
\begin{array}{ccc}
5 & \xi(t) & \zeta(t) \\
\bar \xi(t) & 5+\sqrt{5} & \xi(t) \\
\bar \zeta(t) & \bar \xi(t) & 5
\end{array}
\right)\label{rhomatr}
\ee
with
\be
\xi(t)
&=&
\frac{\left(2+3i-\sqrt{5}i\right)\sqrt{3+\sqrt{5}}}
{\sqrt{3}(e^{\gamma (t-t_0)}+e^{-\gamma (t-t_0)})}
e^{i\omega t},
\\
\zeta(t)
&=&
-
\frac{9e^{2\gamma (t-t_0)}+1+4\sqrt{5}i}
{3(e^{2\gamma(t-t_0)}+1)}
e^{2i\omega t}
\ee
solves the Lax-von Neumann equation (\ref{F20}) (the readers may cross-check it directly in Wolfram Mathematica).
The parameters are
$\omega=1-\frac{5+\sqrt{5}}{15+\sqrt{5}} s$,
$\gamma=\frac{2}{15+\sqrt{5}} s$, and $t_0\in \mathbb{R}$ is arbitrary.

The associated set of rate equations for
$p_A=p_{12}$, $p_B=p_{13}$, $p_C=p'_{12}$, $p_D=p'_{13}$ reads
\be
\dot p_A
&=&
-p_{C}
\frac{1-p_1}{2}
-
s\frac{1-p_1}{2}p_1
+
\nonumber\\
&\pp=&
+
s\Big(
p_1 p_{A}
- \frac{1-p_1}{2} p_{B}
+ \frac{1-p_1}{2} p_{D}
\Big)
\nonumber\\
&\pp=&
+ s\Big(p_{B}p_{C}- p_{A}p_{D}\Big),
\label{pA}\\
\dot p_B
&=&
2p_1- 2p_{D}
-4s p_{A}p_{C}\label{pB}\\
&\pp=&
+
2s\Big(
-p_1
-\frac{(p_1+1)(1-3p_1)}{2}\Big)
\nonumber\\
&\pp=&
+
2s\Big(
(1-p_1) p_{A}+(1-p_1) p_{C}+p_2 p_{D}
\Big),
\nonumber\\
\dot p_C
&=&
-\frac{1-p_1}{2}+p_{A}
+s\Big(p_{A}p_{B}+p_{C}p_{D}\Big)
\nonumber\\
&\pp=&
+
s\Big(
3 \frac{1-p_1}{2}p_1
-2p_1  p_{A}
-\frac{1-p_1}{2} p_{B}\Big)
\nonumber\\
&\pp=&
-
s\Big(p_1 p_{C}
+\frac{1-p_1}{2} p_{D}
\Big),\label{pC}
\\
\dot p_D
&=&
-2p_1+2p_{B}
+
2s\Big(
p_2p_1
-(1-p_1) p_{A}
\Big)
\nonumber\\
&\pp=&
+
2s\Big(
-p_2 p_{B}
+(1-p_1) p_{C}
\Big)
\nonumber\\
&\pp=&
+
2s\Big(p_{A}^2
- p_{C}^2
\Big).\label{pD}
\ee
For this concrete solution the diagonal elements are constants satisfying $p_1=p_3$, $p_1+p_2+p_3=\Tr w=1$.
Additional constants of motion are given by $\Tr(w^n)$ and $\Tr(A w^n)$, for any natural $n$.
An explicit solution of this complicated auto-catalytic system can be extracted from (\ref{rhomatr}). Simplicity of the equivalent Lax-von Neumann equation (\ref{F20}) as compared to (\ref{pA})--(\ref{pD}) is striking.

Beginning again with the linear case $s=0$ we can interpret the black and green curves at Fig.~4 as resources for the consumers evolving according to the red and yellow curves (compare analogous plots for linear and nonlinear consumers given by \cite{WA 2005}). Increasing $s$ we gradually eliminate activity of the ``red" consumer which approaches a steady state.

\section{Periodicity of environment}

Seasonal changes can be modeled by periodic environment which couples to the two consumers and their resources described before.
Periodicity of the environment, in its simplest version, can be modeled by a linear and conservative Lax-von Neumann dynamics of $\varrho_1$.
In spite of high complication of the system of coupled nonlinear rate equations, let us stress again that we deal here with exact 1-soliton solutions. The reader may extract their explicit forms by means of (\ref{rhomatr}), (\ref{rho_2}), and (\ref{p1,12}).

Figs.~7--11 show solutions of the coupled system
\be
\dot p_{1,12}
&=&
k \Big(p_{1,12}'-\frac{1}{2}\Big),\nonumber\\
\dot p_{1,12}'
&=&
-k \Big(p_{1,12}-\frac{1}{2}\Big),\nonumber\\
\dot p_A
&=&
\Bigg[
-p_{C}
\frac{1-p_1}{2}
-
s\frac{1-p_1}{2}p_1
+
\nonumber\\
&\pp=&
+
s\Big(
p_1 p_{A}
- \frac{1-p_1}{2} p_{B}
+ \frac{1-p_1}{2} p_{D}
\Big)
\nonumber\\
&\pp=&
+ s\Big(p_{B}p_{C}- p_{A}p_{D}\Big)\Bigg](\alpha+\beta p_{1,12}),
\nonumber\\
\dot p_B
&=&
(\alpha+\beta p_{1,12})\Bigg[2p_1- 2p_{D}
-4s p_{A}p_{C}\nonumber\\
&\pp=&
+
2s\Big(
-p_1
-\frac{(p_1+1)(1-3p_1)}{2}\Big)
\nonumber\\
&\pp=&
+
2s\Big(
(1-p_1) p_{A}+(1-p_1) p_{C}+p_2 p_{D}
\Big)\Bigg],
\nonumber\\
\dot p_C
&=&
\Bigg[
-\frac{1-p_1}{2}+p_{A}
+s\Big(p_{A}p_{B}+p_{C}p_{D}\Big)
\nonumber\\
&\pp=&
+
s\Big(
3 \frac{1-p_1}{2}p_1
-2p_1  p_{A}
-\frac{1-p_1}{2} p_{B}\Big)
\nonumber\\
&\pp=&
-
s\Big(p_1 p_{C}
+\frac{1-p_1}{2} p_{D}
\Big)\Bigg](\alpha+\beta p_{1,12}),
\nonumber\\
\dot p_D
&=&
\Bigg[-2p_1+2p_{B}
+
2s\Big(
p_2p_1
-(1-p_1) p_{A}
\Big)
\nonumber\\
&\pp=&
+
2s\Big(
-p_2 p_{B}
+(1-p_1) p_{C}
\Big)
\nonumber\\
&\pp=&
+
2s\Big(p_{A}^2
- p_{C}^2
\Big)\Bigg](\alpha+\beta p_{1,12}).
\nonumber
\ee
Initial conditions for probabilities shown in Figs.~5--11 are identical in all the figures.
We can say that the system is uncoupled from its environment if $\alpha=1$, $\beta=0$. Parameters corresponding to $\alpha+\beta/2=0$ lead to periodic dynamics of the system.
Fig.~7 shows the effect of periodic ``hibernation" of populations that can occur in seasons of small
\be
\alpha+\beta p_{1,12}
=
\alpha+\frac{\beta}{2}+\frac{\beta}{2}\cos kt.
\ee
Figs.~7--8 involve strong coupling with environment, and two slightly different values of $s$. Parameter $t_0$ is set to 0 in all these plots, but we have the additional possibility of performing a controlled modification of initial conditions, if we modify $t_0$. Note that this would not be equivalent to just shifting the plots by $t_0$. All density matrices differing by the value of $t_0$ belong to the same ``symplectic leaf" of the dynamics, i.e. they posses the same eigenvalues.

Fig.~12 shows sensitivity of the dynamics with respect to small changes of initial conditions. We again consider the case of no coupling to environment ($\alpha=1$, $\beta=0$, $s=s_{\rm c}+0.1$). The initial conditions correspond to $t_0=30$ (solid curves) and $t_0=100$ (dashed curves). The differences at $t=0$ are smaller than thickness of the plots.

\section{Non-normalized and decaying solutions}\label{decay}

It is clear that the standard context of population dynamics does not require kinetic variables to be interpretable in terms of {\it probabilities\/} (positivity is sufficient). Moreover, typical rate equations involve mortality and birth rate terms that are absent in equations we have discussed so far.

Let us briefly discuss these two issues.
We begin with
\be
i\dot w(t)
&=&
(1-s)[A,w(t)]+s[A,w(t)^2],
\ee
and define a new density operator
\be
v(t)
&=&
e^{i(1-s)At}w(t)e^{-i(1-s)At}
\ee
satisfying
\be
i\dot v(t)
&=&
s[A,v(t)^2].
\ee
Now let
\be
u(t)
&=&
x(t)v\big(y(t)\big)
\ee
where $x(t)$ and $y(t)$ are arbitrary differentiable functions. Then
\be
\dot u(t)
&=&
\frac{\dot x(t)}{x(t)}u(t)
+
\frac{\dot y(t)}{x(t)}(-i)[sA,u(t)^2].
\ee
Finally, the unnormalized but positive matrix
\be
\varrho(t)
&=&
e^{-i(1-s)At}u(t)e^{i(1-s)At}
\nonumber\\
&=&
x(t)e^{i(1-s)A(y(t)-t)}w\big(y(t)\big)e^{-i(1-s)A(y(t)-t)}
\ee
satisfies
\be
i\dot\varrho
=
(1-s)[A,\varrho]
+
\frac{\dot y}{x}s[A,\varrho^2]
+
i\frac{\dot x}{x}\varrho.
\ee
If we are interested in rate equations with no explicit time dependence of parameters, we have to assume that
\be
x(t) = x_0 e^{-\mu t},\quad
y(t) = y_0 e^{-\mu t},
\ee
leading to
\be
\dot\varrho
=
-i(1-s)[A,\varrho]
+
i\frac{\mu y_0s}{x_0}[A,\varrho^2]
-
\mu\varrho.
\ee
Choosing, for example, $x_0=-\mu y_0$, we obtain the equation we have studied before, but with a term describing death (or birth) process,
\be
i\dot\varrho
=
(1-s)[A,\varrho]
+
s[A,\varrho^2]
-
i\mu\varrho,
\ee
which is equivalent to
\be
\dot p_X=- \mu p_X + \dots ,\label{mu}
\ee
$X=A,B,C,D$ and the dots stand for the right-hand-sides of (\ref{pA})--(\ref{pD}).

The next three figures show solutions of (\ref{mu}) for three different couplings with environment. In all these figures we have the same initial conditions, $k=1$, $t_0=0$, $\mu=0.1$, $s=1$, $x_0=1$, $y_0=-x_0/\mu$. In Fig.~13 the coupling parameters are $\alpha=1$, $\beta=0$, meaning that the system is unaffected by its environment. In Fig.~14 $\alpha=1$ but $\beta=-2$. Finally, in Fig.~15 $\alpha=1$ and $\beta=-1.9$.

As we can see, system that would exponentially decay in the absence of interaction with environment, may survive longer and even eliminate the decay if the coupling with environment is strong enough.

It seems an appropriate place to mention the problem of paradox of the plankton \citep{Hutchinson(1961)} or, more generally, the issue of biodiversity. The phenomenon of species oscillations combined with nonlinear feedbacks \citep{HW99} is a candidate for theoretical explanation of the failure of competitive  exclusion principle \citep{gause1935}. Here, periodically changing environment induces a similar phenomenon: The decay is either slowed by a kind of hibernation or even completely blocked by interaction with environment.

So far we have included only the simplest form of dissipation, where all the right-hand sides of the equations are modified by the same linear term $\mu\rho$. One can further elaborate the theory by including the so called Lindblad terms of the form $\Lambda\rho=2M\rho M^\dag-M^\dag M\rho -\rho M^\dag M$ for some linear operator $M$. An appropriately formulated dissipative generalization of von Neumann type equations will be a necessary step in transition from dynamics to thermodynamics of ecosystems \citep{Rodriguez}.

\section{von Neumann form of generalized replicator equations}

Evolutionary game theory \citep{MS,HS} concentrates on evolution of probabilities of strategies. But any game involves two types of probabilities, existing at two essentially different levels. These second probabilities are implicitly present in matrix elements of payoff matrices. The latter probabilities do not couple to probabilities of strategies and play a role of parameters that determine the game. This is what happens at least in standard games, like poker. However, in games where players try to manipulate also the packs of cards, by adding aces hidden up their sleeves, the two levels get mixed with one another.
Replicator equations apply to standard games, but aces-up-one's-sleeves games must involve a generalized formalism. Let us stress that the generalization we have in mind should not be confused with Khrennikov's quantum-like games played by partly irrational players \citep{Khrennikov2012}, or the related field of non-Kolmogorovian aspects of decision making \citep{busemeyerbruza2012}. Our generalization effectively reduces to adding a coupling between formal structures that already exist in the standard formalism.

In \citep{Aerts et al.(2013)} we show that the ``pack of cards" probabilities are in general non-Kolmogorovian. One can say that each entry of a payoff matrix involves probabilities associated with a separate probability space. An equation that mixes the two levels has to be based on a formalism that goes beyond Kolmogorovian probability. The Lax-von Neumann form turns out to be especially useful.

The two types of probabilities enter the standard Kolmogorovian replicator equation,
\be
\frac{dx_k}{dt} &=& x_k\Big(\sum_{l=1}^na_{kl}x_l-\sum_{l,m=1}^na_{lm}x_lx_m\Big),\label{rep eq}\\
a_{lm} &=& \sum_{j=1}^N b_jp_{lm;j},\quad l,m=1,\dots, n; \quad j=1,\dots, N,\nonumber\\
\sum_{k=1}^nx_k &=&1;\quad \sum_{j=1}^Np_{lm;j}=1;\quad x_k\geq 0, p_{lm;j}\geq 0,\nonumber
\ee
in different places. The non-Kolmogorovian ones are denoted by $p_{lm;j}$. The indices $l,m$ in $p_{lm;j}$ index probability spaces in a probability manifold. $p_{lm;j}$ are the ``pack-of-cards" probabilities implicitly present in any game, but treated as fixed and independent of the probabilities of strategies $x_k$. The fact that the replicator equation possesses an equivalent Lax-von Neumann form was discovered by \cite{Pryk}. Their form reads
\be
i\frac{d\rho_1}{dt} &=& [H_1(\rho_1),\rho_1],\nonumber\\
\rho_1 &=& |1\rangle\langle 1|,\nonumber\\
\langle 1| &=& (\sqrt{x_1},\dots,\sqrt{x_n}),\quad x_k=\Tr \rho_1 P_{1,k},\nonumber\\
H_1(\rho_1)&=&i [D_1(\rho_1),\rho_1],\quad D_1(\rho_1)\nonumber\\
&=&\frac{1}{2}{\rm diag}\Big(\sum_{l=1}^na_{1l}x_l,\dots,\sum_{l=1}^na_{nl}x_l\Big).\nonumber
\ee
Here $P_{1,k}={\rm diag}(0,\dots,0,1,0,\dots 0)$, with 1 in the $k$th position.

In order to generalize the Gafiychuk-Prykarpatsky equation to games with aces up one's sleeves one proceeds as follows.
First denote $p_{11,j}=p_j$, $\langle \tilde 1|=(\sqrt{p_1},\dots,\sqrt{p_N})$, and $\rho_2 = |\tilde 1\rangle\langle \tilde 1|$. It can be shown that there exists $E_{lm;j}$ such that $p_{lm;j}=\Tr \rho_2 E_{lm;j}$. $E_{lm;j}=U_{lm}P_{2,j}U_{lm}^\dag$ for some projectors $P_{2,j}$, $p_j=\Tr\rho_2 P_{2,j}$. Denote by $\otimes$ the tensor product, and let $\rho=\rho_1\otimes \rho_2$, $\hat a_{lm}=\sum_j b_j \mathbb{I}_1\otimes E_{lm;j}$. Now, $x_k=\Tr \rho (P_{1,k}\otimes \mathbb{I}_2)$, $p_{lm;j}=\Tr \rho (\mathbb{I}_1\otimes E_{lm;j})$, $a_{lm}=\Tr \rho \hat a_{lm}$.
Denoting
\be
D(\rho) &=& \frac{1}{2}{\rm diag}\Big(\sum_{l=1}^na_{1l}\Tr \rho (P_{1,l}\otimes \mathbb{I}_2),\dots,\sum_{l=1}^na_{nl}\Tr \rho (P_{1,l}\otimes \mathbb{I}_2)\Big)\otimes \mathbb{I}_2,\nonumber\\
H(\rho)&=&i [D(\rho),\rho],\nonumber
\ee
we reconstruct the standard replicator equation by taking the partial trace over the second subsystem from both sides of
\be
i\frac{d\rho}{dt} &=& [H(\rho),\rho],\label{gen rep}
\ee
under the constraints $\rho=\rho_1\otimes \rho_2$, $d\rho_2/dt=0$. The constraints imply $[D(\rho),\rho]= [D_1(\rho_1),\rho_1]\otimes \rho_2$, $d\rho/dt=d\rho_1/dt\otimes \rho_2$. Relaxing the constraints in (\ref{gen rep}), one generalizes (\ref{rep eq}) to games involving players with an ace up their sleeve, where correlations between $x_k$ and $p_{lm;j}$ are no longer ignored.

The question if replicator equations can be reformulated as a soliton system is an open one.

\section{Final remarks}

We have tried to show the potential inherent in soliton von Neumann equations. We have concentrated on the rate-equation aspect of the dynamics. However, if matrices occurring in $H$ are replaced by differential operators one arrives at dynamical variables of a density type. The von Neumann equations then turn into a kind of reaction-diffusion systems, in general of an integro-differential type \citep{Aerts et al.(2003)}. The issues of dissipative systems, such as those preliminarily discussed in Section~\ref{decay}, require more detailed studies. Yet another possibility of including dissipation (based on a complex-time continuation of real-time soliton solutions) can be found in \citep{Aerts Czachor(2007)}. The ace-up-one'e-sleeve generalization of replicator equation will be applied to a certain class of lizard-population evolutionary games in a forthcoming work. In the accompanying paper \citep{Aerts et al.(2013)} we apply some of the ideas discussed here to a real life game played by {\it Uta stansburiana\/} lizards.

\section*{Acknowledgments}

This work was supported by the Flemish Fund for Scientific Research (FWO) projects G.0234.08 and G.0405.08.

\begin{figure}
\includegraphics[width=6cm]{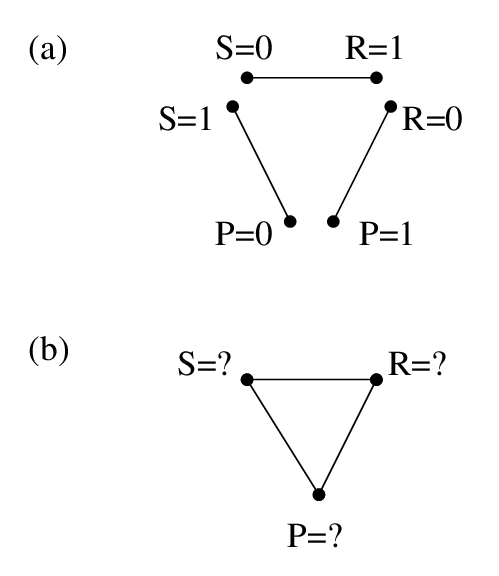}
\caption{Logical and probabilistic non-classicality of the rock-paper-scissors game. (a) There is no problem with three independent games. Results $(S=0,R=1)$, $(P=0,S=1)$, $(R=0,P=1)$, occur with certainty. (b) Simultaneous game with three players is internally inconsistent.}
\end{figure}
\begin{figure}
\includegraphics[width=10cm]{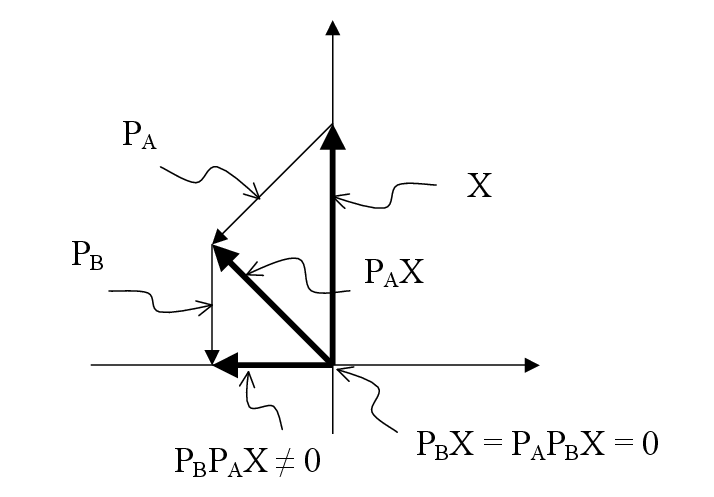}
\caption{Generic non-commutativity of ``questions" represented by projections of vectors. The first ``question" creates a {\it context\/} for the second one. The same ``question" leads to context-dependent probabilities of answers.}
\end{figure}
\begin{figure}
\includegraphics[width=10cm]{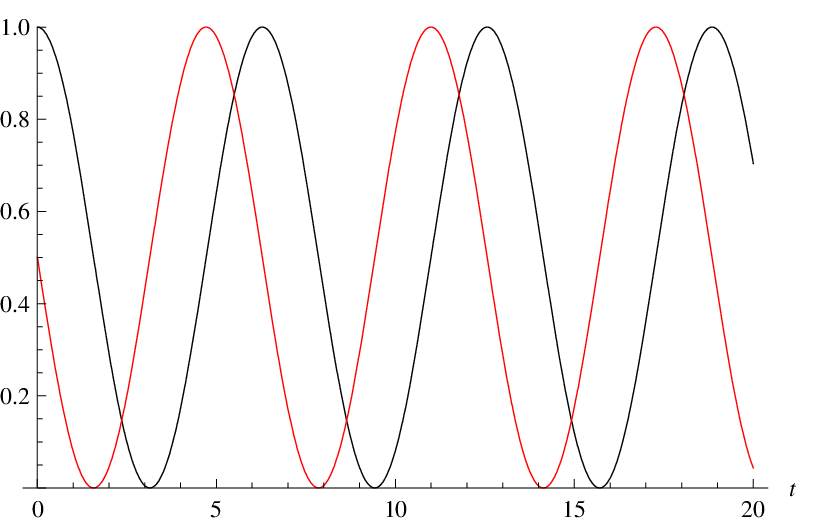}
\caption{Linear model of population $\tilde C$ (black) feeding on the resource $\tilde D$ (red).}
\end{figure}
\begin{figure}
\includegraphics[width=10cm]{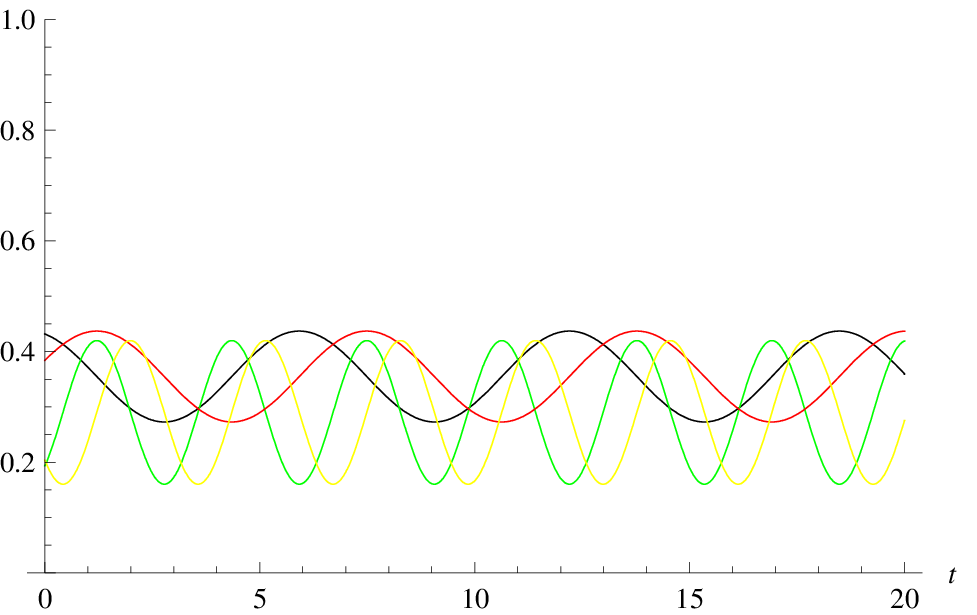}
\caption{Dynamics (\ref{pA})--(\ref{pD}) corresponding to $s=0$, $t_0=0$. We observe two independent consumers $p_C$ and $p_D$ feeding on independent resources $p_A$ and $p_B$. The environment acts trivially since $\alpha=1$, $\beta=0$. For $s=0$ the dynamics is linear. }
\end{figure}
\begin{figure}
\includegraphics[width=10cm]{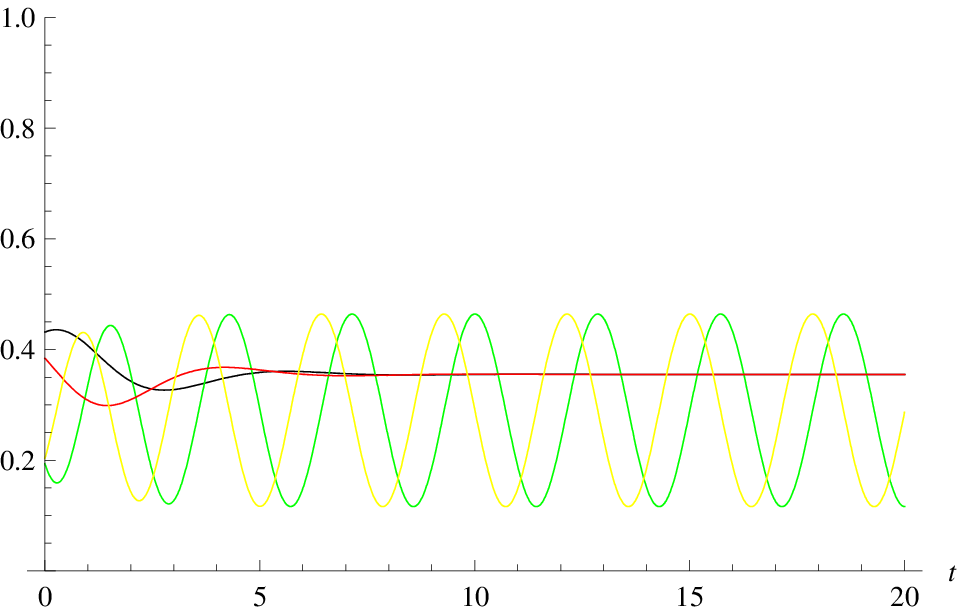}
\caption{Dynamics (\ref{pA})--(\ref{pD}) corresponding to $s=5$ and the same initial condition as in Fig.~4. The environment still acts trivially since $\alpha=1$, $\beta=0$. One of the consumers approaches steady state solution.}
\end{figure}
\begin{figure}
\includegraphics[width=10cm]{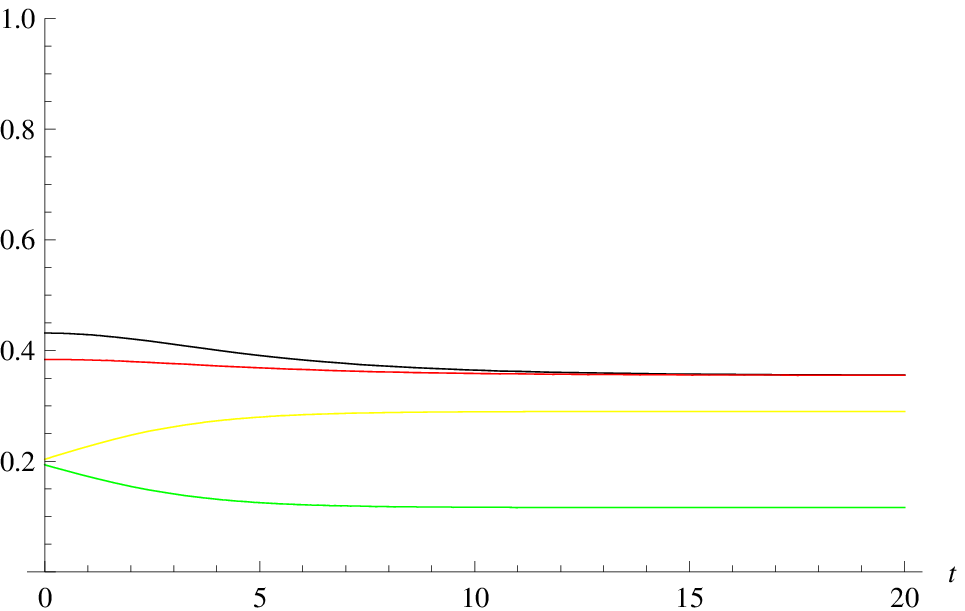}
\caption{Dynamics corresponding to the critical value $s=s_{\rm c}$ for which $\omega=1-\frac{5+\sqrt{5}}{15+\sqrt{5}} s_{\rm c}=0$. The same initial condition as in Fig.~4 and Fig.~5. The environment still acts trivially since $\alpha=1$, $\beta=0$. The system approaches a steady state.}
\end{figure}
\begin{figure}
\includegraphics[width=10cm]{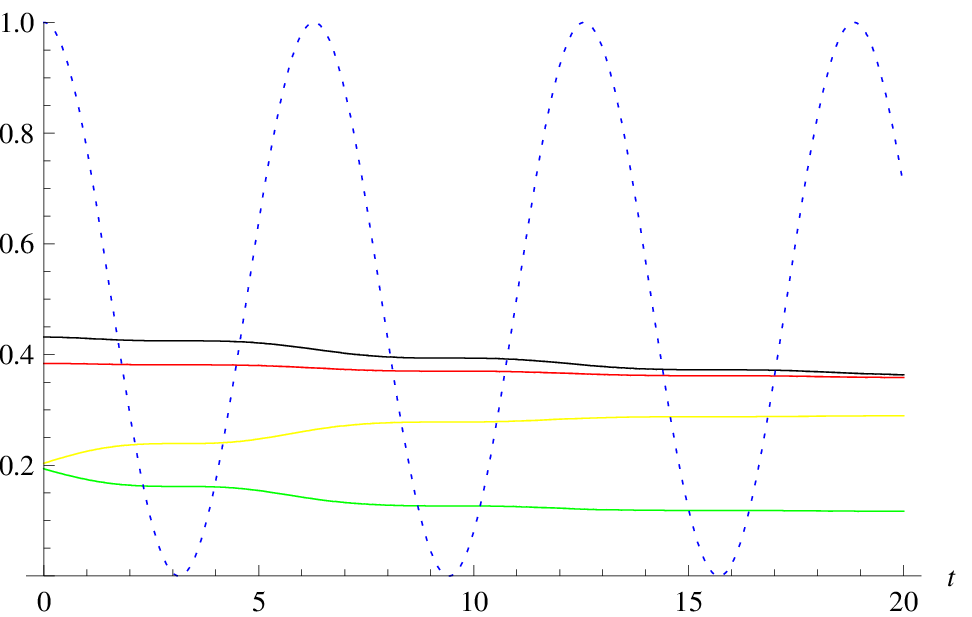}
\caption{The same parameters as in Fig.~6, but now the systems are coupled to an oscillating environment (\ref{p1,12}): $\alpha=0$, $\beta=1$, $k=1$. We observe seasonal ``hibernation" of all the four populations. The dotted curve is the oscillatory probability $p_{1,12}(t)$.}
\end{figure}
\begin{figure}
\includegraphics[width=10cm]{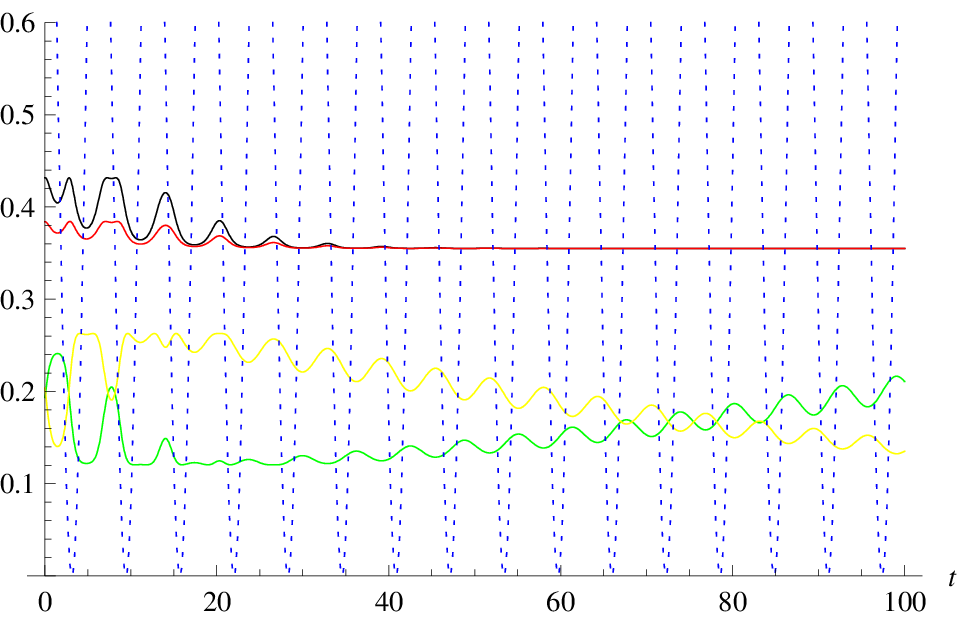}
\caption{The same initial condition as before, but strong coupling to environment $\alpha=5$, $\beta=-9$, $k=1$. $s=s_{\rm c}-0.025$. }
\end{figure}
\begin{figure}
\includegraphics[width=10cm]{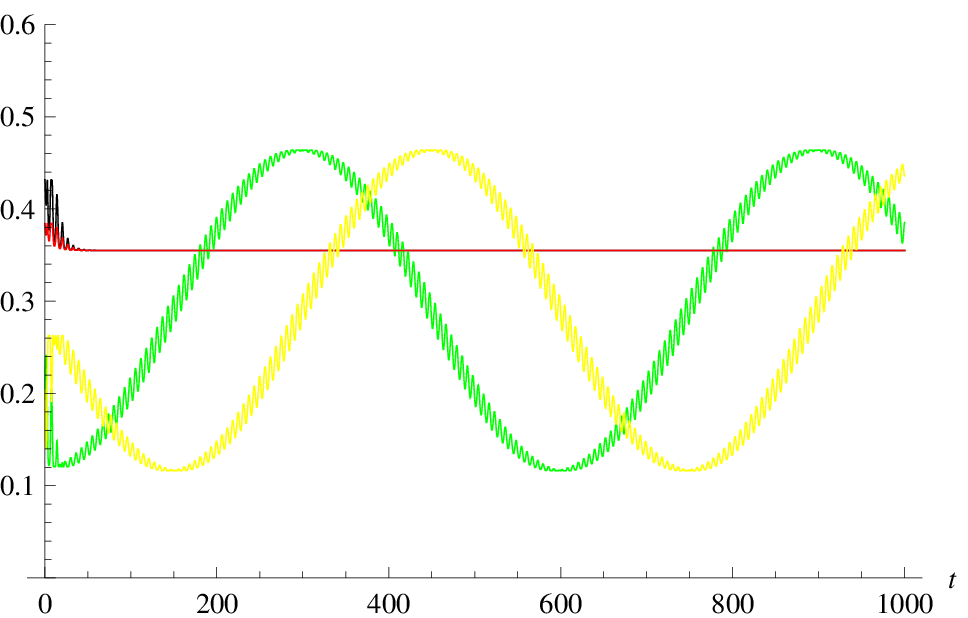}
\caption{The same dynamics as in the previous figure, but for a longer time. We do not show the environment since its oscillation is too rapid in such a time scale. The next plot shows what happens if we increase $s$ by 0.05...}
\end{figure}
\begin{figure}
\includegraphics[width=10cm]{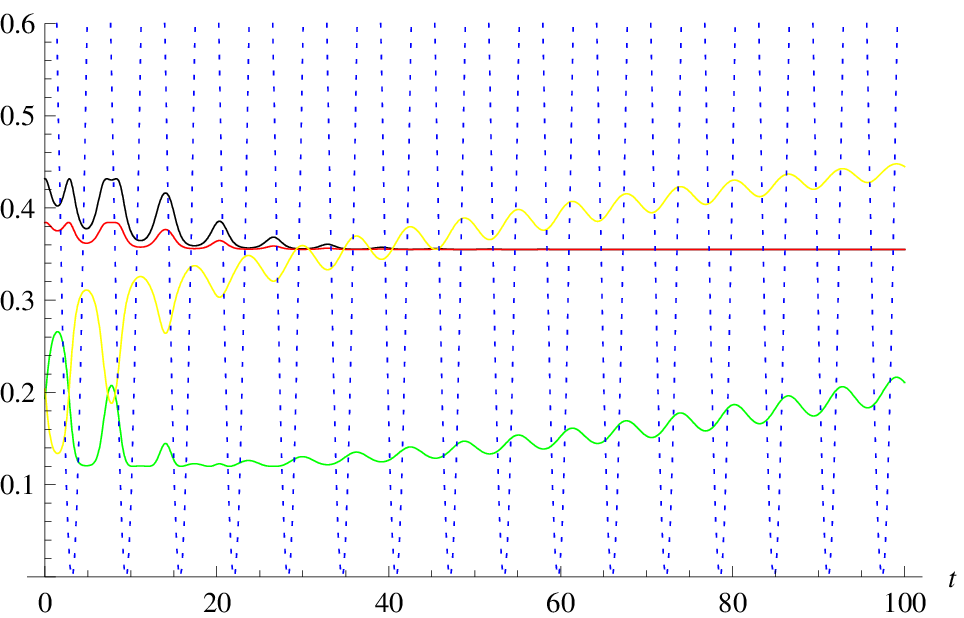}
\caption{The same situation as in the two previous figures, but now $s=s_{\rm c}+0.025$.}
\end{figure}
\begin{figure}
\includegraphics[width=10cm]{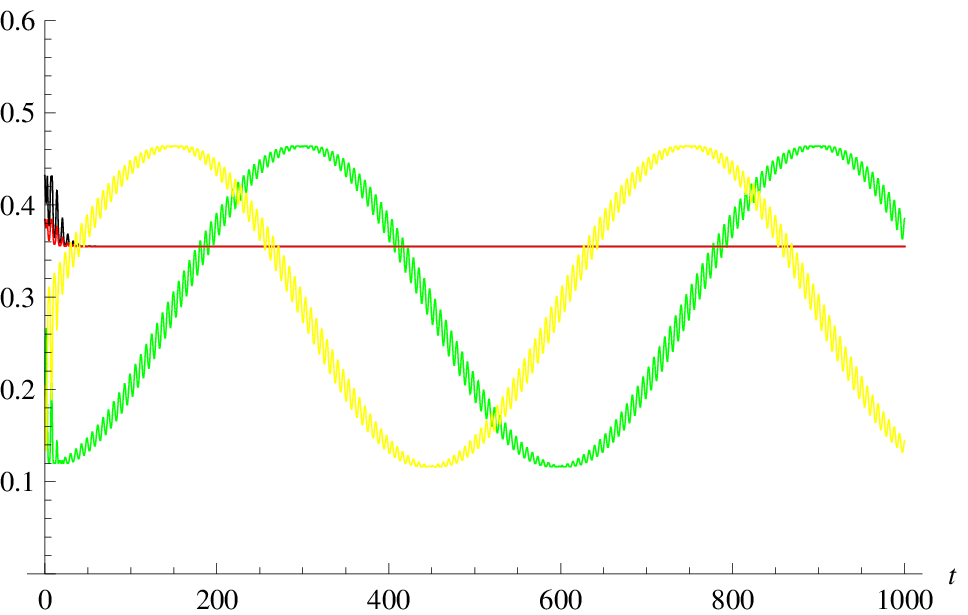}
\caption{A longer-time monitoring of the dynamics from Fig.~10.}
\end{figure}
\begin{figure}
\includegraphics[width=10cm]{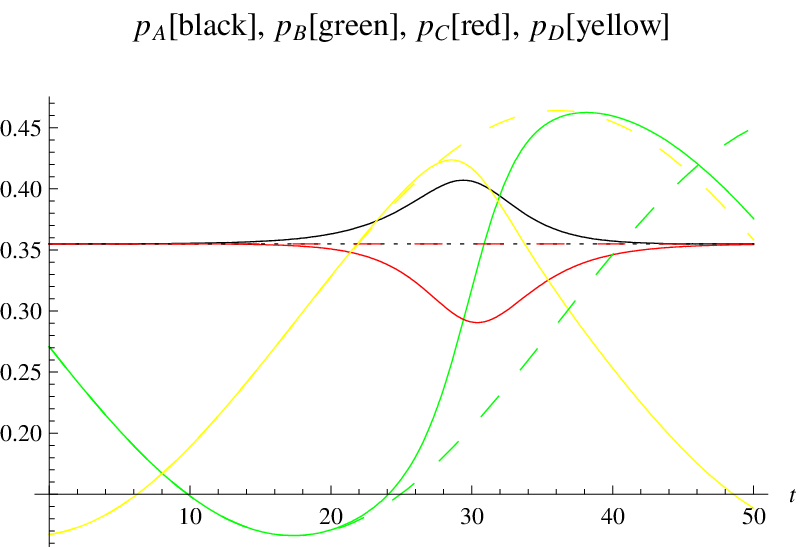}
\caption{The ``butterfly effect". Sensitivity to initial conditions (no coupling to environment). Curves of the same color represent the same kinetic variable, but with slightly different values at $t=0$.}
\end{figure}
\begin{figure}
\includegraphics[width=10cm]{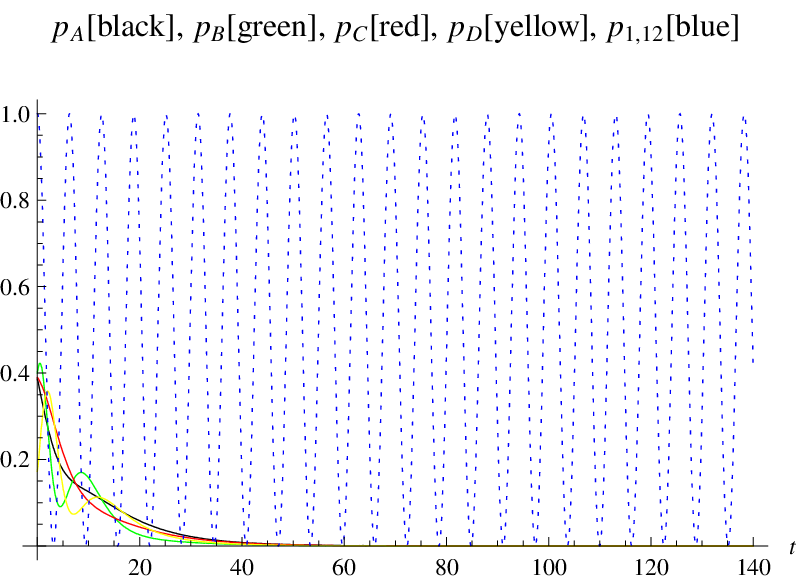}
\caption{Mortality rate is $\mu=0.1$. The remaining parameters are $k=1$, $t_0=0$, $s=1$, $x_0=1$, $y_0=-x_0/\mu$, $\alpha=1$, and $\beta=0$. The latter means that the system is unaffected by its environment}
\end{figure}
\begin{figure}
\includegraphics[width=10cm]{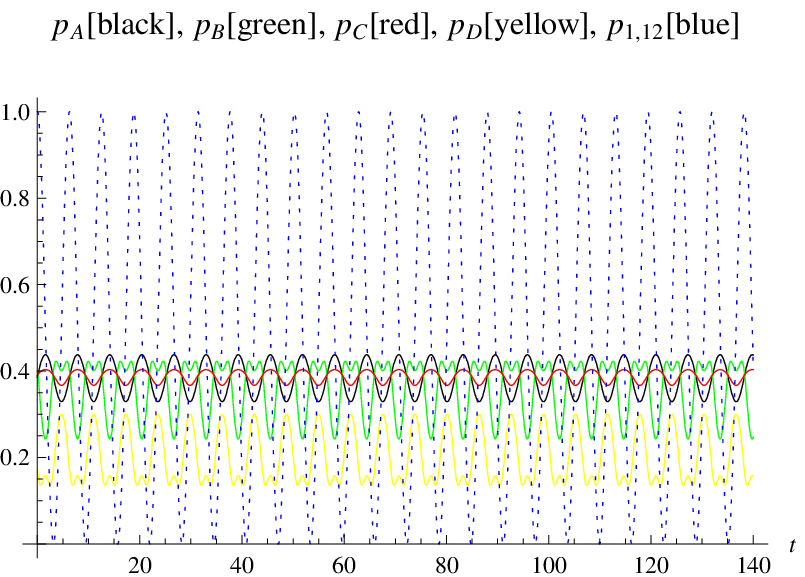}
\caption{The same as in the previous figure, but now $\beta=-2$, so that the environment is coupled to the system. The decay has been stopped due to periodicity of environmental changes. The effect bears some resemblance to the paradox of the plankton. }
\end{figure}
\begin{figure}
\includegraphics[width=10cm]{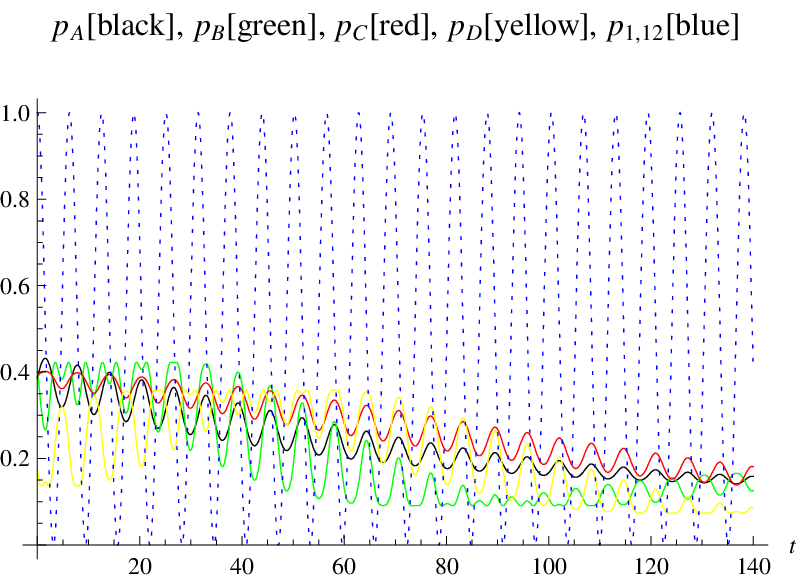}
\caption{The same parameters as in Fig. 14, but with $\beta=-1.9$, i.e. a slightly weaker coupling to environment.}
\end{figure}

\end{document}